\documentclass{ragtime} 

\title[Quasinormal spectrum in the asymptotically safe gravity]%
      {Quasinormal spectrum in the asymptotically safe gravity}
\author[A. F. Zinhailo]
       {A. F. Zinhailo\at[]{1,a}\\
\ins{1}Research Centre for Theoretical Physics and Astrophysics,\splitins[1]
       Institute of Physics, Silesian University in Opava,\splitins[1]                                              
       Bezru\v{c}ovo n\'am.~13, CZ-746\,01 Opava,
       Czech Republic\\
\ins{a}\Email{F170631@fpf.slu.cz}} 

\coentry{A. F. Zinhailo}%
       {Quasinormal spectrum in the asymptotically safe gravity}

\def\imo{i}
\DeclareMathAlphabet{\pazocal}{OMS}{zplm}{m}{n}

\begin{document}

\begin{abstract}
Asymptotically safe gravity is based on the idea of the dependence of the gravitational coupling upon the distance from the origin, approaching its classical value in the weak field regime. We consider three cases of identifying the cut-off parameter in the asymptotically safe gravity, leading to the three distinctive models for black holes. We find that the deviation of the fundamental mode from the Schwarzschild limit is a few percent, in contrast to the higher overtones, where the deviation reaches hundreds of percent, even when the fundamental mode almost coincides with the Schwarzschild mode. This behavior is connected with the fact that the quantum correction to the black hole spacetime is strong near the event horizon, but quickly falls off with distance and negligible near the peak of the effective potential surrounding the black hole.
\end{abstract}


\section{Introduction}\label{intro}
The investigation of quasinormal modes (QNMs) \cite{Kokkotas:1999bd,Nollert:1999ji,Konoplya:2011qq} of black holes has become crucial in comprehending how black holes respond to perturbations. QNMs, representing the characteristic damped oscillations of black holes, encapsulate vital information about their fundamental properties. Their observation is facilitated by gravitational interferometer systems like LIGO and Virgo \cite{LIGOScientific:2016aoc}. Simultaneously, the concept of asymptotically safe gravity \cite{Bonanno:2000ep} has emerged as a theoretical framework aimed at addressing issues regarding the renormalization of gravity in the quantum domain.

This report explores the relationship between QNMs of black holes and the hypothesis of asymptotically safe gravity. We conduct an extensive examination of quasinormal modes of test fields for three black-hole models within asymptotically safe gravity, demonstrating that the overtones are considerably more sensitive to quantum corrections than the fundamental mode. This phenomenon, termed 'the outburst of overtones,' is associated with the distortion of black hole geometry near its event horizon \cite{Konoplya:2022pbc}. This report provides an overview of the primary findings concerning quasinormal spectra of black holes in asymptotically safe gravity. Further details can be explored in \cite{Konoplya:2022hll,Konoplya:2023aph}.

\section{Black hole metrics and wavelike equations}\label{sec.Bardeen spacetime and the wavelike equations}

The metric of a spherically-symmetric black hole has the following form
\begin{equation}\label{spherical}
\mathrm{d}s^2 = -f(r)\mathrm{d} t^2 + f(r)^{-1} \mathrm{d} r^2 + r^2 \: \mathrm{d}\Omega^2.
\end{equation}
In asymptotically safe gravity, Newton's coupling now depends on $r$.
The basic constituent of the theory is the cut-off parameter. 
There are three known ways to identify the cut-off parameter of the theory:
\begin{enumerate}
\item Identification of the cut-off parameter as a modified proper distance leads to the \emph{Bonanno-Reuter metric} \cite{Bonanno:2000ep}.
The metric function for the Bonanno-Reuter spacetime is
\begin{equation}
f(r) = 1-\frac{2 M r^2}{r^3 + \frac{118}{15 \pi} \left(r + \frac{9}{2} M \right)},
\end{equation}
where $M$ is the black hole mass measured in units of the Planck mass.
\item Identification of the cut-off parameter as a function of the Kretschmann scalar \cite{Held:2019xde} leads to the metric coinciding with the \emph{Hayward metric} \cite{Hayward:2005gi}:
\begin{equation}
f(r) = 1-\frac{2 r^2/M^2}{r^{3}/M^{3}+ \gamma}.
\end{equation}
Here the event horizon exists whenever $\gamma \leq 32/27$.

\item  Starting from a classical Schwarzschild solution, the backreaction effects produced by the running Newton's coupling are taken into account iteratively. This way a kind of coordinate independent iterative procedure for identification suggested in \cite{Platania:2019kyx} leads to the Dymnikova black hole \cite{Dymnikova:1992ux}
\begin{equation}\label{f(r)}
f(r) = 1-\frac{2 M}{r} \left(1-e^{-\dfrac{r^3}{2 l_{cr}^2 M}}\right).
\end{equation}
Here $l_{cr}$ is a critical length-scale below which the modifications owing to the running of the Newton’s constant become negligible.
The maximal value of $l_{cr}$ at which the event horizon still exists is
$$l_{cr} \approx 1.138 M,$$
where $M$ is the mass measured in units of length.
\end{enumerate}
The general relativistic equations for the scalar ($\Phi$), electromagnetic ($A_\mu$), and Dirac ($\Upsilon$) fields in a curved spacetime can be cast to the wave-like form:
\begin{equation}\label{wave-equation}
\dfrac{d^2 \Psi}{dr_*^2}+(\omega^2-V(r))\Psi=0,
\end{equation}
where the ``tortoise coordinate'' $r_*$ is:
\begin{equation}\label{tortoise}
dr_*\equiv\frac{dr}{f(r)}.
\end{equation}

The effective potentials for the scalar ($s=0$) and electromagnetic ($s=1$) fields can be written in a unified form:
\begin{equation}\label{potentialScalar}
V(r)=f(r) \frac{\ell(\ell+1)}{r^2}+\left(1-s\right)\cdot\frac{f(r)}{r}\frac{d f(r)}{dr},
\end{equation}
where $\ell=s, s+1, s+2, \ldots$ are the multipole numbers.
For the Dirac field ($s=1/2$) the problem is reduced to two iso-spectral effective potentials
\begin{equation}
V_{\pm}(r) = W^2\pm\frac{dW}{dr_*}, \quad W\equiv \left(\ell+\frac{1}{2}\right)\frac{\sqrt{f(r)}}{r}.
\end{equation}
The isospectral wave functions  $\Psi_{\pm}$ can be transformed one into another by using the Darboux transformation:
\begin{equation}\label{psi}
\Psi_{+}=q \left(W+\dfrac{d}{dr_*}\right) \Psi_{-}, \quad q=const.
\end{equation}
Therefore, it is sufficient to study only one of the iso-spectral cases.

\section{Quasinormal modes}\label{geneqcond}

Quasinormal modes represent the appropriate oscillation frequencies that govern the evolution of perturbations during the intermediate to late stages, known as the ringdown phase. These modes correspond to solutions of the master wave equation (\ref{wave-equation}) subject to the following boundary conditions for the wave function $\Psi\propto {e^{\imo\omega(r_*-t)}}$:
\begin{equation}\label{BC}
\begin{array}{ccll}
    \Psi \sim pure~outgoing~wave, \quad r_* \rightarrow +\infty, \\
    \Psi \sim pure~ingoing~wave, \quad r_* \rightarrow -\infty. \\
\end{array}%
\end{equation}
For finding dominant quasinormal modes, we used the 6th order WKB method  \cite{Konoplya:2003ii,Konoplya:2019hlu} with the Pade approximants \cite{Matyjasek:2017psv}. The WKB method has been broadly applied for finding quasinormal modes of black holes and wormholes (for examples in \cite{Kodama:2009bf,Onozawa:1995vu,Konoplya:2019hlu,Bolokhov:2023ruj}). In order to find precise values of modes with arbitrary relation between the multipole number and overtone, the Leaver method is used \cite{Leaver:1985ax}, which is based on the convergent procedure. Another convergent technique we used is the Bernstein polynomial method \cite{Konoplya:2022zav}, though there only several first overtones can be found during reasonable computing time. 
Finally, in order to see the evolution of perturbations in time, the time-domain integration developed in \cite{Gundlach:1993tp} has been applied. This method was also used in great number of  publications (see for example, \cite{Konoplya:2007yy,Churilova:2019cyt,Bronnikov:2019sbx} and references therein) with a very good concordance with other methods for the fundamental mode.  

\subsection{Bonanno-Reuter metric}

\begin{table}
\begin{tabular}{|l|c|c|}
\hline
  $n$ & $\omega$ (Bonanno-Reuter) & $\omega$ (Schwarzschild) \\
\hline
  $0$ & $0.066877-0.020549\imo$ & $0.062066-0.023122\imo$ \\
  $1$ & $0.059764-0.063499\imo$ & $0.053629-0.073417\imo$ \\
  $2$ & $0.047392-0.110619\imo$ & $0.043693-0.131297\imo$ \\
  $3$ & $0.030448-0.166115\imo$ & $0.036544-0.192977\imo$ \\
  $4$ & $0.019951-0.224263\imo$ & $0.031639-0.255638\imo$ \\
  $5$ & $0.012209-0.290487\imo$ & $0.028063-0.318481\imo$ \\
  $6$ & $0.000983-0.354117\imo$ & $0.025304-0.381317\imo$ \\
  $7$ & $0.000-0.420\imo$       & $0.023081-0.444100\imo$ \\
\hline
\end{tabular}
\caption{Dominant quasinormal modes for the electromagnetic perturbations ($\ell=1$) of the Bonanno-Reuter black hole ($M=4$) and the corresponding modes for the Schwarzschild black hole, according to \cite{Konoplya:2022hll}.}\label{table2e}
\end{table}

Using the first order WKB formula we can obtain quasinormal modes in the high multipole number $\ell\to\infty$ regime in analytic form. For this we will use the expression for the position of the peak of the effective potential, which for the Bonanno-Reuter metric is located at:
\begin{equation}
r_{max} \approx 3 M-\frac{1652}{135 \pi M} -\frac{5071817}{54675 \pi ^2 M^3} 
-\frac{28261793432}{22143375 \pi ^3 M^5}.
\end{equation}
Then, the WKB formula yields
\begin{eqnarray}
Im (\omega) &=& \frac{\left(n+\frac{1}{2} \right)}{3 \sqrt{3} M} \left(1-\frac{1298}{405 \pi M^2} + \mathcal{O}\left(\frac{1}{M^{4}}\right) \right), \\
Re (\omega) &=& \frac{\left(\ell+\frac{1}{2}\right)}{3 \sqrt{3} M} \left(1+\frac{59}{27 \pi M^2} + \mathcal{O}\left(\frac{1}{M^{4}}\right) \right).
\end{eqnarray}

Observing table \ref{table2e}, it becomes evident that as $n$ increases, the deviation of the overtones amplifies, while the fundamental mode only exhibits a slight deviation from the Schwarzschild limit. Additionally, at $n=7$, a purely imaginary (non-oscillatory) mode emerges in the spectrum, which is not an algebraically special one. 

\subsection{Hayward metric}

In a similar fashion for the Hayward metric we obtain the following expressions for the location of the peak of the effective potential,
\begin{equation}
r_{max} \approx 3 M -\frac{2 \gamma M}{9} -\frac{\gamma ^2 M}{27} -\frac{70 \gamma ^3 M}{6561} -\frac{665 \gamma ^4 M}{177147},
\end{equation}
and the eikonal quasinormal frequencies,
\begin{eqnarray}
Im (\omega) &=& \frac{\left(n+\frac{1}{2} \right)}{3 \sqrt{3} M} \left(1-\frac{2}{27} \gamma + \mathcal{O}(\gamma^2) \right), \\
Re (\omega) &=& \frac{\left(\ell+\frac{1}{2} \right)}{3 \sqrt{3} M} \left(1-\frac{2}{54} \gamma + \mathcal{O}(\gamma^2) \right).
\end{eqnarray}
According to observations in \cite{Cardoso:2008bp}, there exists a correspondence between the eikonal quasinormal modes and characteristics of null geodesics: The real and imaginary parts of $\omega$ are multiples of the frequency and instability timescale of the circular null geodesics, respectively. While we confirm this observation for the specific case considered in asymptotically safe gravity, generally, this correspondence does not hold for modes with $\ell \gg n$ that cannot be accurately reproduced by the WKB formula \cite{Konoplya:2017wot,Konoplya:2022gjp}. 

Table \ref{table5} shows that there is an outburst of overtones for the Hayward metric as well. 

\begin{tiny}
\begin{table}
\begin{tabular}{|l|c|c|}
\hline
$n$ & $\omega M$ (Hayward) & $\omega M$ (Schwarzschild)  \\
\hline
  $0$ & $0.113494-0.089160 \imo$ & $0.110455-0.104896\imo$ \\
  $1$ & $0.066731-0.319873\imo$ & $0.086117-0.348053\imo$ \\
  $2$ & $0.041068-0.576924\imo$ & $0.075742-0.601079\imo$ \\
  $3$ & $0.021679-0.833067\imo$ & $0.070410-0.853678\imo$\\
  $4$ & $0.000000-1.082236\imo$ & $0.067074-1.105630\imo$ \\
  $5$ & $0.001449-1.317232\imo$ & $0.064742-1.357140\imo$ \\
\hline
\end{tabular}
\caption{Quasinormal modes for the $\ell=0$ scalar perturbations of the Hayward black hole ($\gamma=1$) and the corresponding modes for the Schwarzschild black hole, according to \cite{Konoplya:2022hll}.}\label{table5}
\end{table}
\end{tiny}

\subsection{Dymnikova black hole}

As can be seen from Table \ref{tabl:Frobenius:s=1} the overtones deviate at a stronger rate from their Schwarzschild values when $n$ is increased.  For all three types of metrics, the deformation of the geometry occurs mainly near the event horizon, while in the far zone, the metrics merge with the Schwarzschild one. The overtones are highly sensitive to these near horizon deformations.  

\begin{tiny}
\begin{table*}
\begin{tabular}{|l|c|c|c|c|}
  \hline
  $l_{cr}$            & $n=0$                 & $n=1$               & $n=2$             & $n=3$ \\
  \hline
  $0$ & $0.24826-0.09249\imo$ & $0.2145-0.2937\imo$ & $0.175-0.525\imo$ & $0.146-0.772\imo$ \\ 
  $0.7$               & $0.24823-0.09247\imo$ & $0.2141-0.2935\imo$ & $0.173-0.525\imo$ & $0.140-0.772\imo$ \\ 
  $0.75$              & $0.24814-0.09239\imo$ & $0.2135-0.2930\imo$ & $0.170-0.523\imo$ & $0.135-0.768\imo$ \\ 
  $0.8$               & $0.24804-0.09226\imo$ & $0.2125-0.2922\imo$ & $0.166-0.521\imo$ & $0.123-0.765\imo$ \\ 
  $0.85$              & $0.24790-0.09200\imo$ & $0.2112-0.2909\imo$ & $0.160-0.518\imo$ & $0.103-0.762\imo$ \\ 
  $0.9$               & $0.24766-0.09159\imo$ & $0.2088-0.2886\imo$ & $0.149-0.512\imo$ & $0.06~-0.74~\imo$ \\ 
  $0.95$              & $0.2473~-0.09085\imo$ & $0.206~-0.2849\imo$ & $0.13~-0.50~\imo$ & $0.06~-0.81~\imo$ \\ 
  $1.0$               & $0.2468~-0.08986\imo$ & $0.201~-0.280~\imo$ & $0.107-0.508\imo$ & $0.05~-0.8~~\imo$ \\ 
  $1.05$              & $0.24614-0.08855\imo$ & $0.195~-0.276~\imo$ & $0.107-0.514\imo$ & $0.04~-0.8~~\imo$ \\ 
  $1.1$               & $0.24516-0.08699\imo$ & $0.1892-0.2731\imo$ & $0.091-0.519\imo$ & $0.05~-0.87~\imo$ \\ 
   \hline
\end{tabular}
\caption{Quasinormal modes found by the Leaver method for $\ell=1$, electromagnetic perturbations; $M=1$. The metric is approximated by the 17th-order parametrization in \cite{Konoplya:2023aph}. The Schwarzschild limit corresponds to $l_{cr} =0$.}\label{tabl:Frobenius:s=1}
\end{table*}
\end{tiny}

\section{Conclusions}\label{conclus}

We have reviewed recent results obtained in \cite{Konoplya:2023aph,Konoplya:2022hll} regarding the behavior of overtones in various black hole models within asymptotically safe gravity. Despite different approaches in identifying the cutoff parameters, a qualitatively similar feature is observed in all three cases: an outburst of overtones that convey information about the geometry of the event horizon.

\ack

I would like to thank Roman Konoplya for the most useful discussions. 

\bibliography{ragsamp}

\begin{thebibliography}{27}
\expandafter\ifx\csname natexlab\endcsname\relax\def\natexlab#1{#1}\fi
\expandafter\ifx\csname url\endcsname\relax
  \def\url#1{\texttt{#1}}\fi
\expandafter\ifx\csname urlprefix\endcsname\relax\def\urlprefix{URL }\fi
\providecommand{\selectlanguage}[1]{\relax}
\providecommand{\eprint}[2][]{\url{#2}}

\bibitem[{Abbott et~al.(2016)}]{LIGOScientific:2016aoc}
Abbott, B.~P. et~al. (LIGO Scientific, Virgo) (2016), {Observation of Gravitational Waves from a Binary Black Hole Merger}, \emph{Phys. Rev. Lett.}, \textbf{116}(6), p. 061102, \eprint{1602.03837}.

\bibitem[{Bolokhov(2023)}]{Bolokhov:2023ruj}
Bolokhov, S.~V. (2023), {Long lived quasinormal modes and telling oscillatory tails of the Bardeen spacetime, 10.20944/preprints202310.0517.v1}.

\bibitem[{Bonanno and Reuter(2000)}]{Bonanno:2000ep}
Bonanno, A. and Reuter, M. (2000), {Renormalization group improved black hole space-times}, \emph{Phys. Rev.}, \textbf{D62}, p. 043008, \eprint{hep-th/0002196}.

\bibitem[{Bronnikov and Konoplya(2020)}]{Bronnikov:2019sbx}
Bronnikov, K.~A. and Konoplya, R.~A. (2020), {Echoes in brane worlds: ringing at a black hole--wormhole transition}, \emph{Phys. Rev. D}, \textbf{101}(6), p. 064004, \eprint{1912.05315}.

\bibitem[{Cardoso et~al.(2009)Cardoso, Miranda, Berti, Witek and Zanchin}]{Cardoso:2008bp}
Cardoso, V., Miranda, A.~S., Berti, E., Witek, H. and Zanchin, V.~T. (2009), {Geodesic stability, Lyapunov exponents and quasinormal modes}, \emph{Phys. Rev. D}, \textbf{79}(6), p. 064016, \eprint{0812.1806}.

\bibitem[{Churilova and Stuchlik(2020)}]{Churilova:2019cyt}
Churilova, M.~S. and Stuchlik, Z. (2020), {Ringing of the regular black-hole/wormhole transition}, \emph{Class. Quant. Grav.}, \textbf{37}(7), p. 075014, \eprint{1911.11823}.

\bibitem[{Dymnikova(1992)}]{Dymnikova:1992ux}
Dymnikova, I. (1992), {Vacuum nonsingular black hole}, \emph{Gen. Rel. Grav.}, \textbf{24}, pp. 235--242.

\bibitem[{Gundlach et~al.(1994)Gundlach, Price and Pullin}]{Gundlach:1993tp}
Gundlach, C., Price, R.~H. and Pullin, J. (1994), {Late time behavior of stellar collapse and explosions: 1. Linearized perturbations}, \emph{Phys. Rev. D}, \textbf{49}, pp. 883--889, \eprint{gr-qc/9307009}.

\bibitem[{Hayward(2006)}]{Hayward:2005gi}
Hayward, S.~A. (2006), {Formation and evaporation of regular black holes}, \emph{Phys. Rev. Lett.}, \textbf{96}, p. 031103, \eprint{gr-qc/0506126}.

\bibitem[{Held et~al.(2019)Held, Gold and Eichhorn}]{Held:2019xde}
Held, A., Gold, R. and Eichhorn, A. (2019), {Asymptotic safety casts its shadow}, \emph{JCAP}, \textbf{06}, p. 029, \eprint{1904.07133}.

\bibitem[{Kodama et~al.(2010)Kodama, Konoplya and Zhidenko}]{Kodama:2009bf}
Kodama, H., Konoplya, R.~A. and Zhidenko, A. (2010), {Gravitational stability of simply rotating Myers-Perry black holes: Tensorial perturbations}, \emph{Phys. Rev. D}, \textbf{81}, p. 044007, \eprint{0904.2154}.

\bibitem[{Kokkotas and Schmidt(1999)}]{Kokkotas:1999bd}
Kokkotas, K.~D. and Schmidt, B.~G. (1999), {Quasinormal modes of stars and black holes}, \emph{Living Rev. Rel.}, \textbf{2}, p.~2, \eprint{gr-qc/9909058}.

\bibitem[{Konoplya(2003)}]{Konoplya:2003ii}
Konoplya, R.~A. (2003), {Quasinormal behavior of the d-dimensional Schwarzschild black hole and higher order WKB approach}, \emph{Phys. Rev. D}, \textbf{68}, p. 024018, \eprint{gr-qc/0303052}.

\bibitem[{Konoplya(2023)}]{Konoplya:2022gjp}
Konoplya, R.~A. (2023), {Further clarification on quasinormal modes/circular null geodesics correspondence}, \emph{Phys. Lett. B}, \textbf{838}, p. 137674, \eprint{2210.08373}.

\bibitem[{Konoplya and Fontana(2008)}]{Konoplya:2007yy}
Konoplya, R.~A. and Fontana, R. D.~B. (2008), {Quasinormal modes of black holes immersed in a strong magnetic field}, \emph{Phys. Lett. B}, \textbf{659}, pp. 375--379, \eprint{0707.1156}.

\bibitem[{Konoplya and Stuchl\'\i{}k(2017)}]{Konoplya:2017wot}
Konoplya, R.~A. and Stuchl\'\i{}k, Z. (2017), {Are eikonal quasinormal modes linked to the unstable circular null geodesics?}, \emph{Phys. Lett. B}, \textbf{771}, pp. 597--602, \eprint{1705.05928}.

\bibitem[{Konoplya et~al.(2023)Konoplya, Stuchlik, Zhidenko and Zinhailo}]{Konoplya:2023aph}
Konoplya, R.~A., Stuchlik, Z., Zhidenko, A. and Zinhailo, A.~F. (2023), {Quasinormal modes of renormalization group improved Dymnikova regular black holes}, \emph{Phys. Rev. D}, \textbf{107}(10), p. 104050, \eprint{2303.01987}.

\bibitem[{Konoplya and Zhidenko(2011)}]{Konoplya:2011qq}
Konoplya, R.~A. and Zhidenko, A. (2011), {Quasinormal modes of black holes: From astrophysics to string theory}, \emph{Rev. Mod. Phys.}, \textbf{83}, pp. 793--836, \eprint{1102.4014}.

\bibitem[{Konoplya and Zhidenko(2022)}]{Konoplya:2022pbc}
Konoplya, R.~A. and Zhidenko, A. (2022), {First few overtones probe the event horizon geometry}, \eprint{2209.00679}.

\bibitem[{Konoplya and Zhidenko(2023)}]{Konoplya:2022zav}
Konoplya, R.~A. and Zhidenko, A. (2023), {Bernstein spectral method for quasinormal modes of a generic black hole spacetime and application to instability of dilaton\textendash{}de Sitter solution}, \emph{Phys. Rev. D}, \textbf{107}(4), p. 044009, \eprint{2211.02997}.

\bibitem[{Konoplya et~al.(2019)Konoplya, Zhidenko and Zinhailo}]{Konoplya:2019hlu}
Konoplya, R.~A., Zhidenko, A. and Zinhailo, A.~F. (2019), {Higher order WKB formula for quasinormal modes and grey-body factors: recipes for quick and accurate calculations}, \emph{Class. Quant. Grav.}, \textbf{36}, p. 155002, \eprint{1904.10333}.

\bibitem[{Konoplya et~al.(2022)Konoplya, Zinhailo, Kunz, Stuchlik and Zhidenko}]{Konoplya:2022hll}
Konoplya, R.~A., Zinhailo, A.~F., Kunz, J., Stuchlik, Z. and Zhidenko, A. (2022), {Quasinormal ringing of regular black holes in asymptotically safe gravity: the importance of overtones}, \emph{JCAP}, \textbf{10}, p. 091, \eprint{2206.14714}.

\bibitem[{Leaver(1985)}]{Leaver:1985ax}
Leaver, E.~W. (1985), {An Analytic representation for the quasi normal modes of Kerr black holes}, \emph{Proc. Roy. Soc. Lond. A}, \textbf{402}, pp. 285--298.

\bibitem[{Matyjasek and Opala(2017)}]{Matyjasek:2017psv}
Matyjasek, J. and Opala, M. (2017), {Quasinormal modes of black holes. The improved semianalytic approach}, \emph{Phys. Rev. D}, \textbf{96}(2), p. 024011, \eprint{1704.00361}.

\bibitem[{Nollert(1999)}]{Nollert:1999ji}
Nollert, H.-P. (1999), {TOPICAL REVIEW: Quasinormal modes: the characteristic `sound' of black holes and neutron stars}, \emph{Class. Quant. Grav.}, \textbf{16}, pp. R159--R216.

\bibitem[{Onozawa et~al.(1996)Onozawa, Mishima, Okamura and Ishihara}]{Onozawa:1995vu}
Onozawa, H., Mishima, T., Okamura, T. and Ishihara, H. (1996), {Quasinormal modes of maximally charged black holes}, \emph{Phys. Rev. D}, \textbf{53}, pp. 7033--7040, \eprint{gr-qc/9603021}.

\bibitem[{Platania(2019)}]{Platania:2019kyx}
Platania, A. (2019), {Dynamical renormalization of black-hole spacetimes}, \emph{Eur. Phys. J. C}, \textbf{79}(6), p. 470, \eprint{1903.10411}.

\end{thebibliography}

\end{document}